# Online Ordering Platform City Distribution Based on Genetic Algorithm

DU Yu

School of Economics and Management, Beijing Jiaotong University, China, 100044

**Abstract:** Since the rising of the takeaway ordering platform, the M platform has taken the lead in the industry with its high-quality service. The increasing order volume leads the competition between platforms to reduce the distribution cost, which increases rapidly because of the unreasonable distribution route. By analyzing platform distribution's current situation, we study the vehicle routing problem of urban distribution on the M platform and minimize the distribution cost. Considering the constraints of the customer's expected delivery time and vehicle condition, we combine the different arrival times of the vehicle routing problem model using three soft time windows and solve the problem using a genetic algorithm (GA). The results show that our model and algorithm can design the vehicle path superior to the original model in terms of distribution cost and delivery time, thus providing decision support for the M platform to save distribution cost in urban distribution in the future.

## 1 Introduction

When customers make an order, the terminal of the delivery station allocates the order according to the order quantity and the location. The courier selects the driving route according to his experience. When some places are gathering together in a one-time trip, which causes delay, and other separate assignments in remote areas result in higher distribution costs. Therefore, the M platform can only optimize the delivery cost and reduce the uncertainty in meal delivery by continuously optimizing the delivery route of the delivery truck.

The practical problem of the M platform in urban distribution is a kind of vehicle routing problem (VRP). The delivery of takeaway is two steps: taking the meal and delivering the meal. The courier starts from a random point and goes to the merchant to take the meal and then completes the delivery within the prescribed time. Our paper mainly studies the meal delivery problem of the food delivery staff. In addition, the customers will have different requirements for the arrival time of the food delivery. Thus, it is necessary to introduce a time window. We will use a specific Shanghai area to study the M-online ordering platform to adopt the VRP with the time window (VRPTW). We use the appropriate algorithm to work out the path planning model and change the way of selecting the path according to experience, minimizing the distribution cost.

Through the research on the urban distribution of M online ordering platform, we propose a new urban distribution VRP to reduce the current peak delivery time, change the chaos and delay, save the platform operation cost and improve the platform revenue.

## 2 Literature review

The service scope of the M platform is within the city, so the distribution process belongs to urban distribution. Urban distribution refers to the logistics activities for urban and suburban services. In economically reasonable areas, according to customers' needs, deliver the appropriate goods to customers timely. Urban distribution is the last mile and is the direct communication channel between logistics companies and consumers. Urban distribution is the final step of the logistics chain. With the systematic development of logistics activities, the research is mainly on the distribution mode and distribution cost.

Li [1] started from the discussion of distribution demand in the era of urban development, analyzed the advantages and disadvantages of the traditional and crowdsourcing delivery model, and proposed the urban distribution model of conventional express crowdsourcing. Yang [2] researched the last-mile delivery of e-commerce and built an integrated delivery model that combines public electronic pick-up, manual pick-up points, and direct delivery to avoid or reduce delivery failures, optimizing resource allocation. Zhang [3] proposed distributing mixed low-temperature goods to provide the lowest price, highest efficiency, best service, and timely delivery to each store. Konur [4] analyzed integrated inventory control and distribution scheduling with economic and environmental considerations in a stochastic demand environment. Finally, Teodor [5] combined the Internet of Things with the urban distribution system to propose a super-connected urban distribution network model.

With increased customer expectations and the emergence of numerous customer-oriented businesses, the consideration of delivery time and distribution efficiency in the urban distribution has attracted much attention—our model targets delivery time and total costs, including time expenses. In the actual distribution process, the conventional factors such as the change of traffic lights, urban road restrictions, and unexpected characteristics such as road congestion, road maintenance, and vehicle damage can cause uncertainty in driving time. Frizzell [6] first studied the Split Delivery VRPTW (SDVRPTW) and used a local search algorithm to solve and optimized the two local search algorithms. Yan Fang [7] based on maximum customer satisfaction and minimum total distribution cost, established a mathematical model of multi-fuzzy VRPTW. He used the particle swarm optimization algorithm to solve the model. However, there is little research about the random batch distribution VRP. Yang [8] studied SDVRPTW, established an opportunity constrained programming model, and designed a genetic algorithm. Russell [9] constructed a multi-objective model and developed the tabu search algorithm based on the number of vehicles, total driving distance, and penalty cost.

In summary, the research of urban distribution model provides a way of thinking for the study of the urban distribution of M online ordering platform. Our focus is the optimization of urban vehicle routing. Therefore, we consider the waiting situation under the soft time window, optimize the urban distribution route of the M online ordering platform to shorten the delivery time and reduce the distribution cost.

We intend to solve the problem of urban distribution route planning for the M online ordering platform. We combined the actual distribution process under the M platform's delivery mode, taking Shanghai as an example, analyzing the existing data obtained by the Tianchi Big Data platform for the overall ordering period of 11:30-18:30. Both 11:30 and 13:00 are meal peak periods, combined with the regional distribution of merchants and customers, using the time-segment allocation order method, solving the VRPSTW to study the urban distribution plan of the M online ordering platform. The constraints are vehicle load. We establish an optimization model with the minimum transportation cost, the time penalty cost, and the distance penalty cost. Besides, the time penalty cost function is a three-stage time window function. We use a genetic algorithm to solve the problem, and finally, compare the benefits with the old urban distribution plan.

## 3 Problem Formulation

### 3.1 Factors affecting order delivery

We mainly consider the distribution cost, delivery time, vehicle volume. We choose the VRP model with the minimum distribution cost. The customer will be willing to consider the delivery time. We set a three-stage time window function to calculate the time cost.

### 3.2 Model assumptions

We made the following assumptions to convert the VRPSTW into a mathematical model.

1) Each vehicle shall not deliver more than its maximum load when performing the delivery task.

2) Each node customer only be served once.

3) Each order needs to be taken once and can't be split.

4) The courier starts from any position and does not need to return to the merchant after completing the delivery.

5) The courier's driving path avoids generating sub-loops.

### 3.3 Penalty function

We consider two penalty functions. One is the distance penalty cost, and the other is the time penalty cost. When the courier exceeds the distance, the additional cost $c_L$ will be generated. If the delivery service violates the customer time window requirement, there will be a cost $\beta_i$. As the figure shows, the time window $[a, b]$ meets the delivery time required by the customer. The distribution between $[0, a]$ and $[b, +\infty]$ will result in a corresponding penalty cost.

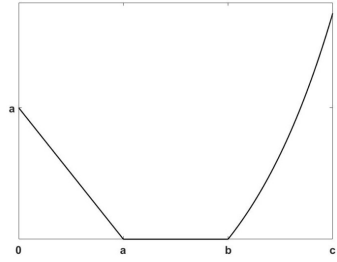

**Fig 1 Time penalty cost function**

The expression of the time penalty cost function is as follows:

$$penalty\ \beta_i = \begin{cases} 0.5 \times (-t_i + a) & 0 \leq t_i < a \\ 0 & a \leq t_i \leq b \\ 1.5 \times (e^{t_i - b} - 1) & b < t_i \leq c \end{cases} \quad (3-1)$$

The parameters in the formula are defined as follows:

$t_i$: the actual arrival time of the courier

$a$: the earliest arrival time the customer expects

$b$: the latest arrival time the customer expects

When the courier arrives early, the customer dissatisfaction will decrease as the lead time decreases, using a linear subtraction function. When the courier comes within the customer's expected time, the customer's dissatisfaction and the time penalty cost are approximately zero. When the courier is late, since the waiting time increases, the meal may not be fresh. We use the exponential function for penalty cost.

The expression for the penalty function is as follows:

$$\alpha_k = \left\lfloor \frac{Total\ travel\ distance\ of\ k\ vehicle}{Total\ endurance\ distance\ of\ k\ vehicle} \right\rfloor \quad (3-2)$$

However, we don't consider the battery capacity. Thus, the courier can work all the time. If the total distance is longer than the theoretical distance, the number is rounded down. For example, 0.5, the penalty cost function is 0. If it is equal to 2.2, Then the penalty cost function is 2, and multiplying the penalty amount is the penalty cost.

### 3.4 Parameter definition

Based on the above description of the problem and the basic assumptions, the variables and parameter symbols are defined as follows:

**Table 1 Params and variables**

| Notation | |
|---|---|
| $V = \{1,2,\ldots,v\}$ | set of vehicles |
| $W = \{1,2,\ldots,w\}$ | set of restaurants |
| $C = \{1,2,\ldots,c\}$ | set of customers |
| $D_i$ | the demand for each order $i$ |
| $d_{ij}$ | the distance between order $i$ and $j$, $i \neq j \in (W \cup C)$ |
| $q$ | per product weight |
| $o$ | per distance cost |
| $r$ | fixed cost per car |
| $t_{mi}$ | the number of products of merchant $m$ each time order $i$ |
| $L$ | The maximum transportation distance of a single battery per unit vehicle |
| $Q$ | the maximum load for the unit vehicle |
| $c_L$ | distance penalty cost |
| $\alpha_k$ | vehicle $k$ distance penalty cost factor |
| $\beta_i$ | order $i$ time penalty cost |
| $f_{ki}$ | When vehicle $k$ finishes order $i$ |
| $x_{ijk}$ | when vehicle $k$ is from customer $i$ to customer $j$ |
| $y_{ijk}$ | when vehicle $k$ is from the merchant $i$ to the customer $j$ or from the customer $i$ to the merchant $j$ |

## 4 Vehicle routing problem model with soft time window

We aim to minimize the distribution cost of the platform by reducing the delivery path and satisfying the needs of all customers. Considering the distribution of customer nodes, service time window, vehicle maximum distance, and real-time traffic information. We construct the following model of VRPSTW.

$$\min Z = \sum_{k \in V}\left[ o\left(\sum_{i \in C}\sum_{j \in C}d_{ij}x_{ijk} + \sum_{i \in W}\sum_{j \in C}d_{ij}y_{ijk} + \sum_{i \in C}\sum_{j \in W}d_{ij}y_{ijk}\right)\right] + vr + \sum_{k \in V}\alpha_k c_L + \sum_{i \in C}\beta_i \quad (4-1)$$

$$T_{mi} = D_i, \quad \forall i \in C, m \in W \quad (4-2)$$

$$\sum_{k \in V}\left[\sum_{i \in C \cap i \neq j}x_{ijk} + \sum_{m \in W}y_{mjk}\right] = 1, \quad \forall j \in C \quad (4-3)$$

$$\sum_{j \in C}\sum_{k \in V}\left[\sum_{i \in C \cap i \neq j}x_{ijk} + \sum_{m \in W}y_{mjk}\right] = c \quad (4-4)$$

$$\sum_{m \in W} \sum_{i \in C} f_{ki} t_{mi} q \leq Q \qquad (4-5)$$

A brief description of the various formulas in the model is as follow:

The model's objective function (4-1) is mainly composed of four parts: transportation distance cost, vehicle fixed cost, distance penalty cost, and penalty time of soft time window. The transportation distance cost is the driving distance and the cost per unit distance. The time cost is formula (3-1), and the distance penalty cost is the penalty amount that multiplies the distance penalty cost formula (3-2). Formula (4-2) indicates that the order quantity received by the dispatcher each time is the sum of the orders. Formula (4-3) guarantees that each order can only be delivered once. Formula (4-4) means that all orders must be delivered. Formula (4-5) is the load constraint of the vehicle.

## 5 Solution Analysis and Numerical Studies

Use Python to write the codes, and the results are as follows:

**Table 2 Model calculation results**

| shop_id | fitness | min C | time | Route |
|---|---|---|---|---|
| S451 | 0.00548 | 182.577 | 28.689 | 0 - 1 - 2 - 4 – 3 |
| S455 | 0.00236 | 423.236 | 36.391 | 0 - 2 - 5 - 6 - 1 – 3 |
|  |  |  |  | 0 - 7 – 4 |
| S464 | 0.00215 | 464.650 | 26.902 | 0 - 8 - 4 - 3 - 1 – 7 |
|  |  |  |  | 0 - 5 - 6 – 2 |
| S471 | 0.00676 | 147.951 | 26.465 | 0 - 3 - 2 - 4 – 1 |
| S474 | 0.00447 | 223.662 | 34.320 | 0 - 3 - 1 - 2 – 4 |
| S478 | 0.00787 | 126.986 | 20.898 | 0 - 3 - 1 - 4 – 2 |
| S486 | 0.00192 | 520.553 | 33.880 | 0 - 3 - 4 – 7 |
|  |  |  |  | 0 - 2 - 1 - 5 – 6 |
| S489 | 0.00127 | 785.004 | 39.046 | 0 - 18 - 16 – 14 |
|  |  |  |  | 0 - 8 - 12 - 2 - 21 - 10 – 7 |
|  |  |  |  | 0 - 5 - 1 - 11 – 4 |
|  |  |  |  | 0 - 19 - 3 - 20 - 15 - 13 - 17 – 9 |
|  |  |  |  | 0 – 6 |
| S491 | 0.00314 | 318.679 | 37.386 | 0 - 2 - 5 - 3 - 1 – 4 |
| S493 | 0.01193 | 83.836 | 20.508 | 0 - 2 - 4 - 1 – 3 |
| S495 | 0.00338 | 295.864 | 34.088 | 0 - 11 - 8 – 13 |
|  |  |  |  | 0 - 10 - 9 - 3 - 7 - 4 - 1 - 12 - 5 - 2 – 6 |

The delivery path as shown:

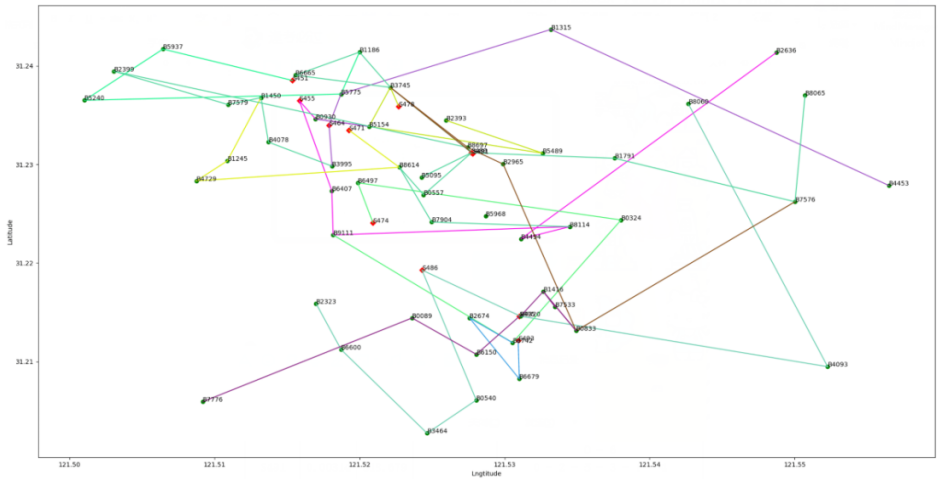

**Fig 2 Distribution route map**

The current shipping cost that according to the objective function, is 3572.999. According to the objective function, the travel time of this round is 338.573. According to the objective function, the driving distance of this round is 74.736.

## 5.1 Program comparison

An example of the verification of the model is to compare with the design adopted by the M online ordering platform at present.

**Table 3 Original program operation result**

| shop_id | fitness | min C | time | route |
|---|---|---|---|---|
| S451 | 0.00548 | 182.577 | 28.689 | 0 - 1 - 2 - 4 - 3 |
| S455 | 0.00215 | 465.713 | 41.784 | 0 - 4 - 1 - 6 - 5 - 2 - 7 - 0 |
| | | | | 0 - 3 - 0 |
| S464 | 0.00250 | 400.329 | 50.433 | 0 - 1 - 5 - 7 - 8 - 4 - 3 - 2 - 6 - 0 |
| S471 | 0.00676 | 147.951 | 26.465 | 0 - 3 - 2 - 4 - 1 |
| S474 | 0.00447 | 223.662 | 34.320 | 0 - 3 - 1 - 2 - 4 |
| S478 | 0.00787 | 126.986 | 20.898 | 0 - 3 - 1 - 4 - 2 |
| S486 | 0.00175 | 570.161 | 42.031 | 0 - 6 - 5 - 4 - 7 - 0 |
| | | | | 0 - 1 - 2 - 3 - 0 |
| S489 | 0.00087 | 1144.089 | 42.122 | 0 - 12 - 19 - 9 - 6 - 5 - 0 |
| | | | | 0 - 16 - 18 - 7 - 11 - 4 - 0 |
| | | | | 0 - 14 - 1 - 8 - 17 - 10 - 0 |
| | | | | 0 - 15 - 20 - 21 - 2 - 3 - 0 |
| | | | | 0 - 13 - 0 |
| S491 | 0.00310 | 322.250 | 45.435 | 0 - 5 - 2 - 4 - 1 - 3 - 0 |
| S493 | 0.01193 | 83.836 | 20.508 | 0 - 2 - 4 - 1 - 3 |
| S495 | 0.00285 | 351.384 | 42.913 | 0 - 8 - 3 - 1 - 5 - 6 - 2 - 11 - 12 - 9 - 0 |
| | | | | 0 - 13 - 4 - 7 - 10 - 0 |

The current objective function is 4018.937. Compared to the original model with the model we proposed from the delivery time and distribution cost, the travel time is 395.598. The travel distance is 97.560.

*5.2 The impact of new and old programs on business benefits*

It can be seen from the above data that the new distribution scheme is more suitable for the long-term development of the enterprise than the original one. Because the M online ordering platform is continuously expanding its business, the order quantity will undoubtedly increase. Our solution is convenient for arranging the route in many orders, saving the distribution cost. At the same time, we consider the distribution time window constraints in line with reality. By increasing the time penalty cost, the platform can consider customers in the distribution, rather than simply pursuing low costs. The development of the platform not only depends on its technology but also the reputation among customers. Only when the courier completes the delivery within the time expected by the customer can the desire of ordering customers be stimulated and increase the number of orders. Thus, the platform will bring more business cooperation and increase platform revenue.

## 6 Conclusion and Future Research

The data shows that the model proposed in this paper is significantly lower than the original model in terms of distribution cost when the order quantity is increasing and the time is shorter. The model also considers the expected time of the customer and the road conditions, which is richer than the original model. We provide support for the M platform in the delivery route selection of daily operating vehicles. We design a VRPSTW for the M ordering platform based on the research on the distribution route of ordering platforms at home and abroad. However, some aspects are not considered very thoughtful, and the algorithm of the model solving part is also more singular. Further research is needed.

# References

[1] L. Qiang, Research on the Construction of Urban Distribution Model System in the Internet Era, Knowledge Economy, (2017) 38-39.
[2] Y. Juping, Y. Changchun, Y. Xuanxia, The Last Mile Problem in E-commerce Logistics, Journal of Business Economics, (2014) 16-22+32.
[3] Z. Mingyu, Y. Chao, W. Shuxiang, W. Wenbing, Research on optimization of mixed-match delivery for chain supermarkets based on resource integration, Business Review, 29 (2017) 223-233.
[4] D. Konur, J.F. Campbell, S.A. Monfared, Economic and environmental considerations in a stochastic inventory control model with order splitting under different delivery schedules among suppliers, Omega, 71 (2017) 46-65.
[5] T.G. Crainic, B. Montreuil, Physical internet enabled hyperconnected city logistics, Transportation Research Procedia, 12 (2016) 383-398.
[6] P.W. Frizzell, J.W. Giffin, The split delivery vehicle scheduling problem with time windows and grid network distances, Computers & Operations Research, 22 (1995) 655-667.
[7] F. Yan, Y. Wang, Modeling and solving the vehicle routing problem with multiple fuzzy time windows, in: International conference on management science and engineering management, Springer, 2017, pp. 847-857.
[8] Y. Xinfeng, Y. Qingfeng, Model and Algorithm of Stochastic Vehicle Routing Problem, Journal of Transportation Systems Engineering and Information Technology, (2006) 75-80.
[9] R.A. Russell, T.L. Urban, Vehicle routing with soft time windows and Erlang travel times, Journal of the Operational Research Society, 59 (2008) 1220-1228.